\documentclass[12pt]{article}
\usepackage{a4wide}

\newcommand{\bbbone}{\hbox{\rm 1\kern-3pt l}}

\newcommand{\Jp}{J/\psi}

\newcommand{\GeV}{\mbox{\,GeV}}

\newcommand{\mub}{\mbox{\,$\mu$b}}
\newcommand{\nb}{\mbox{\,nb}}

\begin{document}

\thispagestyle{empty} 

\begin{center} 
{\Large\bf On the kinematic limit of the charm production\\[-0.5cm]
in fixed-target experiments with the\\
intrinsic charm from the target}\\[1cm]
{\large Stefan Groote, Sergey Koshkarev}\\[0.3cm]
Institute of Physics, University of Tartu, 51010 Tartu, Estonia\\[3pt]
\end{center}

\vspace{1cm}
\begin{abstract}
In this note we provide a detailed derivation of the kinematic limit of the
charm production in fixed-target experiments with the intrinsic charm coming
from the target. In addition, we discuss the first measurement of charm quark
production in the fixed-target configuration at the LHC.
\end{abstract}

\vspace{1cm}
\section{Introduction}

The production of charmed hadrons via the intrinsic charm mechanism from the
target was proposed in Ref.~\cite{Brodsky:2016fyh}\footnote{see chapter~6.1}
and Ref.~\cite{Brodsky2017}. Utilizing this idea, in
Refs.~\cite{Groote:2017szb,Koshkarev:2017txl} production of double charmed
baryons and the $\Jp$ meson from the target was investigated at the STAR and
LHCb/SMOG fixed-target programs~\cite{Meehan:2016iyt,Aaij:2011er}. By a
straightforward calculation it was shown that in case of the production of
intrinsic charm from the target, charm hadrons have a maximum momentum in the
laboratory frame of only a few $\GeV/c$ and a rapidity $y<2$. This result is 
in perfect agreement with predictions from the intrinsic charm model about
the $\Delta y$ region for the charm state,
$\Delta y=y-y_{\rm tar}(y_{\rm tar}=0)<2$.

Recently, the LHCb collaboration presented the first measurement of charm
production in the fixed-target configuration at the LHC~\cite{Aaij:2018ogq}.
The production of $\Jp$ and $D^0$ mesons is studied with beams of protons of
different energies colliding with gaseous targets of helium and argon with
nucleon--nucleon centre-of-mass energies of $\sqrt{s}=86.6$ and $110.4\GeV$,
respectively. The $\Jp$ and $D^0$ production cross-sections in $p$He
collisions in the rapidity range $[2,4.6]$ are found to be
$\sigma(\Jp)=652\pm 33\pm 42\nb$ per nucleon and 
$\sigma(D^0)=80.8\pm 2.4\pm 6.3\mub$ per nucleon, where the first uncertainty
is statistical and the second is systematic. Using an approximation for the
fraction $x$ of the nucleon momentum carried by the target parton,\footnote{We
provide the equation in the form it was used in the LHCb
publication~\cite{Aaij:2018ogq}. However, it is easy to see that this equation
is not fully correct. Using the standard Feynman-$x$ notation in the
center-of-mass system $x\approx\frac{2 m_T}{\sqrt{s}}\sinh(y^*)$ (cf.\
Eq.~(47.43) in Ref.~\cite{Tanabashi:2018oca}) and taking into account that
$\sinh(y^*)\approx-\exp(-y^*)/2$ for large negative values of $y^*$, one
finally obtains $x\approx-\frac{m_T}{\sqrt{s}}\exp(-y^*)$.}
\begin{equation}
x \simeq \frac{2 m_c}{\sqrt{s_{NN}}}\exp(-y^*),
\end{equation}
where $m_c=1.28\GeV/c^2$ is the mass of the $c$ quark and $y^*$ is
centre-of-mass rapidity, Providing an analysis of the production of $D^0$
mesons with the intrinsic charm from the target, LHCb reported no evidence for
a substantial intrinsic charm content of the nucleon~\cite{Aaij:2018ogq}.

In this short note we provide a detailed derivation of the kinematic limit of
the charm production in the fixed-target experiments with the intrinsic charm
coming from the target. In addition, we show that based on a misunderstanding
of the kinematics of the intrinsic charm from the target in the laboratory
framem the LHCb/SMOG result is in contradiction to the properties of the
intrinsic charm model.

\section{Derivation of the kinematic limit}

\subsection{Calculation of the Center of Mass System energy}
Given an invariant energy $\sqrt s$, one can easily calculate the energies in
the laboratory frame, the center-of-mass frame and of course in a lot of
different other frames. Starting from
\begin{equation}
s=(p_P+p_T)^2=p_P^2+2p_Pp_T+p_T^2=m_P^2+2p_Pp_T+m_T^2
\end{equation}
we first can calculate in the laboratory system where $p_T=(m_T;0,0,0)$. One
obtains
\begin{equation}\label{sval}
s=m_P^2+m_T^2+2E_P^{\rm lab}m_T
\end{equation}
For the center-of-mass system, however, we have $\vec p_P+\vec p_T=\vec 0$
($\Rightarrow\vec p_P=-\vec p_T=:\vec p$) and, therefore,
$\sqrt s=E_P+E_T$ where $E_P=(m_P^2+\vec p\,^2)^{1/2}$,
$E_T=(m_T^2+\vec p\,^2)^{1/2}$. Inserting these energies we obtain
\begin{equation}
\sqrt{m_P^2+\vec p\,^2}+\sqrt{m_T^2+\vec p\,^2}=\sqrt s,
\end{equation}
and this equation can be solved by two-fold squaring,
\begin{eqnarray}
2\sqrt{m_P^2+\vec p\,^2}\sqrt{m_T^2+\vec p\,^2}&=&s-m_P^2-m_T^2-2\vec p\,^2
  \nonumber\\
4(m_P^2+\vec p\,^2)(m_T^2+\vec p\,^2)&=&(s-m_P^2-m_T^2-2\vec p\,^2)^2
  \nonumber\\
4\vec p\,^4+4\vec p\,^2(m_P^2+m_T^2)+4m_P^2m_T^2&=&
  (s-m_P^2-m_T^2)^2-4\vec p\,^2(s-m_P^2-m_T^2)+4\vec p\,^4\nonumber\\
4s\vec p\,^2&=&(s-m_P^2-m_T^2)^2-4m_P^2m_T^2\ =\ \lambda(s,m_P^2,m_T^2),\qquad
\end{eqnarray}
where $\lambda(x,y,z)=x^2+y^2+z^2-2xy-2xz-2yz$ is K\"all\'en's function. Given
$s$ by Eq.~(\ref{sval}), one can proceed to calculate the square of the
three-momentum to be
\begin{eqnarray}
\vec p\,^2&=&\frac{\lambda(m_P^2+m_T^2+2E_Pm_T,m_P^2,m_T^2)}{4(m_P^2+m_T^2
  +2E_P^{\rm lab}m_T)}\ =\ \frac{4((E_P^{\rm lab})^2-m_P^2)m_T^2}{4(m_P^2+m_T^2
  +2E_P^{\rm lab}m_T)}\ =\nonumber\\
  &=&\frac{((E_P^{\rm lab})^2-m_P^2)m_T^2}{m_P^2+m_T^2+2E_P^{\rm lab}m_T}
  \ \approx\ \frac{(E_P^{\rm lab})^2m_T^2}{2E_P^{\rm lab}m_T}
  \ =\ \frac12E_P^{\rm lab}m_T.
\end{eqnarray}

\subsection{Lorentz transformations}
In order to get from one system to another, four-vectors have to be multiplied
by Lorentz matrices. The Lorentz matrix which puts the general target
four-vector $p_T=(E_T;-\vec p\,)$ in the CMS to rest is given by
\begin{equation}
\left(\Lambda^\mu{}_\nu(E_T;-\vec p\,)\right)
  =\Lambda(p_T)=\pmatrix{E_T/m_T&\vec p\,^T/m_T\cr
  \vec p/m_T&\bbbone+\vec p\vec p\,^T/m_T/(E_T+m_T)\cr}
\end{equation}
where $m_T^2=E_T^2-\vec p\,^2$. The upper index $T$ stands for transposition
of the (column) three-vector $\vec p\,$. It can easily be seen that
$\Lambda(p_T)p_T=(m;\vec 0)^T$. Assuming that $J/\psi$ takes over a ratio
$x\in[0,1]$ of the three-momentum $\vec p\,$ (for $x>0$ from the projectile,
for $x<0$ from the target) and applying the Lorentz transformation to the
corresponding four-vector $p_\psi=(E_\psi;x\vec p\,)$ with
$E_\psi^2=m_\psi^2+x^2\vec p\,^2$, one obtains a value for the four-vector in
the laboratory frame, i.e.\ the frame where $p_T$ is at rest. The result reads
\begin{eqnarray}
E_\psi^{\rm lab}&=&\frac1{m_T}\left(\sqrt{\vec p_T^{\,2}+m_T^2}
  \sqrt{x^2\vec p_T^{\,2}+m_\psi^2}-x\vec p_T^{\,2}\right),\nonumber\\
\vec p_\psi^{\rm\,lab}&=&\frac{\vec p_T}{m_T}\sqrt{x^2\vec p_T^{\,2}+m_\psi^2}
  -x\vec p_T\left(1+\frac{\vec p_T^{\,2}}{m_T(m_T+\sqrt{\vec p_T^{\,2}+m_T^2})}
  \right).
\end{eqnarray}
Taking into account that the three-momenta are usually much larger than the
masses, we can expand into $m_T^2/\vec p\,^2$ and $m_\psi^2/\vec p\,^2$ to
$x<0$ (target) and obtain the maximum values of energy and momentum 
of the charm in the laboratory frame 
\begin{equation}\label{eq:kin}
E_\psi^{\rm lab}=\frac1{2m_T}\left(m_\psi^2+m_T^2\right),\qquad
\vec p_\psi^{\rm\,lab}=\frac{\vec p}{2m_T|\vec p\,|}
  \left(m_\psi^2-m_T^2\right)
\end{equation}
This last expression depend solely on the two masses $m_\psi$ and $m_T$ and
no longer on the beam energy.

\section{A comment on the LHCb/SMOG result}

Let us remind the reader that in the fixed-target configuration, the LHCb
acceptance gives access to the large Bjorken-$x$ region of the nucleon target
(up to $x\sim 0.37$ for $D^0$ mesons) and to rapidity region $2 < y < 4.6$ in
the laboratory frame. Using Eq.~(\ref{eq:kin}) we can obtain the kinematic
limit (i.e.\ $x_F = 1$) of the charm produced by the intrinsic charm mechanism
from the target in the laboratory frame, $y < 0.6$ and $p_L\sim 1.2-1.3\GeV/c$.
In other words, the LHCb collaboration provided an analysis outside the signal
region.

\section{Conclusion}

In this note we provide a detailed derivation of the kinematic limits of the
charm production in fixed-target experiments with the intrinsic charm coming
from the target. We have also shown that the LHCb/SMOG result is based on a
misunderstanding of the kinematics of the intrinsic charm. Moreover, the
investigation of the production of charmed hadrons with intrinsic charm from
the target at the LHCb/SMOG fixed-target program is fundamentally unfeasible.

\subsection*{Acknowledgements}
We would like to thank Stanley J.~Brodsky for pointing out the LHCb/SMOG
result to us and stimulating discussions. We would also like to thank Ramona
Vogt, Paul Hoyer and Jean-Philippe Lansberg for useful comments. This research
was supported by the Estonian Research Council under Grants No.~TK133
and~PRG356.


\begin{thebibliography}{99}

\bibitem{Brodsky:2016fyh}
  S.~J.~Brodsky, V.~A.~Bednyakov, G.~I.~Lykasov, J.~Smiesko and S.~Tokar,\\
  Prog.\ Part.\ Nucl.\ Phys.\  {\bf 93} (2017) 108

\bibitem{Brodsky2017}
  S.J.~Brodsky, Workshop on LHCb Heavy Ion and Fixed Target physics,\\
  CERN, January 9--10, 2017.

\bibitem{Groote:2017szb}
  S.~Groote and S.~Koshkarev,
  Eur.\ Phys.\ J.\ C {\bf 77} (2017) no.8,  509

\bibitem{Koshkarev:2017txl}
  S.~Koshkarev and S.~Groote,
  J.\ Phys.\ Conf.\ Ser.\  {\bf 938} (2017) no.1,  012054

\bibitem{Meehan:2016iyt}
  K.~C.~Meehan [STAR Collaboration],
  J.\ Phys.\ Conf.\ Ser.\  {\bf 742} (2016) no.1,  012022.
  
\bibitem{Aaij:2011er}
  R.~Aaij {\it et al.} [LHCb Collaboration],
  JINST {\bf 7} (2012) P01010

\bibitem{Aaij:2018ogq}
  R.~Aaij {\it et al.} [LHCb Collaboration],
  [arXiv:1810.07907 [hep-ex]].

\bibitem{Tanabashi:2018oca}
  M.~Tanabashi {\it et al.} [Particle Data Group],
  Phys.\ Rev.\ {\bf D98} (2018) 030001

\end{thebibliography}
\end{document}